\documentclass[aps,prb,showpacs,unsortedaddress]{revtex4}

\usepackage{amsmath}
\usepackage{amssymb}
\usepackage{graphicx}
\usepackage{epsfig}

\begin{document}

\title{A numerical method for generation of quantum noise and solution
of generalized c-number quantum Langevin equation}

\author{Dhruba Banerjee}
\affiliation{Indian Association for the Cultivation of Science,
Jadavpur, Kolkata 700 032, India}

\author{Bidhan Chandra Bag}
\affiliation{Department of Chemistry, Visva Bharati, Shantiniketan 731 235, India}

\author{Suman Kumar Banik}
\affiliation{Indian Association for the Cultivation of Science,
Jadavpur, Kolkata 700 032, India}
\altaffiliation{Present Address:
Max-Planck-Institut f\"ur Physik komplexer Systeme,
N\"othnitzer Str. 38, 01187 Dresden, Germany}

\author{Deb Shankar Ray}
\email{pcdsr@mahendra.iacs.res.in}
\affiliation{Indian Association for the Cultivation of Science,
Jadavpur, Kolkata 700 032, India}

\date{\today}

\begin{abstract}
Based on a coherent state representation of noise operator and an ensemble 
averaging procedure we have recently developed 
[Phys. Rev. E {\bf 65}, 021109 (2002); {\it ibid.} 051106 (2002)]  a scheme for
quantum Brownian motion to derive the equations for time evolution of 
{\it true} probability distribution functions in $c$-number phase space.
We extend the treatment to develop a numerical method for generation
of $c$-number noise with arbitrary correlation and strength at any 
temperature, along with the solution of
the associated generalized quantum Langevin equation.
The method is illustrated with the help of a calculation of quantum
mean first passage time in a cubic potential to demonstrate quantum
Kramers turnover and quantum Arrhenius plot.
\end{abstract}


\maketitle


\section{Introduction}

A system coupled to its environment is the standard paradigm for 
quantum theory
of Brownian motion 
\cite{weiss,whl,ggdsr,grab,woly,mak,makri,dali,pcvl,topa,ank,stoc}. 
Its overwhelming success in the treatment of various
phenomena in quantum optics, transport processes in Josephson junction,
coherence effects and macroscopic quantum tunneling in condensed matter 
physics, electron transfer in large molecules, thermal activation processes 
in chemical reactions is now well known and forms a large body of current 
literature. While the early development of quantum optics initiated 
in the
sixties and seventies was based on density operator, semigroup, noise 
operator or master equation methods primarily within weak-coupling and 
Markov approximations, path integral approach to quantum Brownian motion
attracted wide attention in the early eighties. Although this development 
had widened the scope of condensed matter and chemical physics significantly,
so far as the large coupling between the system and the heat bath and large
correlation times of the noise processes are concerned several problems still
need to be addressed. {\it First}, a search for quantum analogue of Kramers' 
equation for a nonlinear system which describes quantum Brownian motion in
phase space had remained elusive and at best resulted in equation of motion
which contains higher (than second)
derivatives of probability distribution functions \cite{whl,jrc} whose 
positive definiteness is never
guaranteed. As a result these quasi-probability distribution functions often
become singular or negative in the full quantum domain. 
{\it Second}, although long
correlation times and large coupling constants are treated non-perturbatively
formally in an exact manner by path integrals, their analytic evaluation 
is generally completed within semi-classical schemes
\cite{weiss,grab,woly,ank}.
This results in 
situations where the theory fails to retain its validity in the vacuum limit.
{\it Third}, although the numerical techniques 
based on the path integral Monte Carlo
are very successful in the treatment of equilibrium properties it is often
difficult to implement in a dynamical scheme because of the well-known sign
problem associated with varying phases in a quantum evolution over paths 
due to the oscillating nature of the real time propagator
\cite{mak,makri,topa,stoc}. 
{\it Fourth}, the 
treatment of non-Markovian dynamics within a quantum formulation as 
retardation effect in a memory functional is very difficult, if not 
impossible, to handle even numerically \cite{topa,stoc}. 

Keeping in view of these difficulties it is worthwhile to ask how to extend 
classical theory of Brownian motion to quantum domain
for arbitrary friction and temperature.
Based on a coherent state representation of the noise operator and an 
ensemble averaging procedure we have recently \cite{skb,db1,db2,db3} 
developed a scheme for quantum 
Brownian motion in terms of a c-number generalized Langevin equation. The 
approach allows us to use classical methods of non-Markovian dynamics to 
derive exact generalized quantum Kramers' equation and other relevant 
quantum analogues of classical diffusion, Fokker-Planck and Smoluchowski
equations. The object of the present analysis is to extend the treatment 
({\it{i}}) to develop a numerical method for generation of quantum noise as 
classical c-numbers ({\it{ii}}) and to solve the associated generalized
quantum Langevin equation. The numerical method valid for arbitrary 
temperature, damping strength and noise correlation, is exemplified by an 
application to calculation of quantum mean first passage time
in a cubic potential. The approach 
is classical-looking in form, and is independent of path integral quantum
Monte Carlo techniques and takes care of quantum effect order by order with
simplicity and accuracy to a high degree.

The outlay of the paper is as follows: We start with a brief review of 
$c$-number representation of quantum noise and the associated quantum
Langevin equation, as formulated recently by us, in Sec.II. This is 
followed by a numerical method for generation of quantum noise and solution
of the Langevin equation in Sec.III. In Sec.IV we illustrate the procedure by
calculating quantum mean first passage time in a cubic potential with an 
emphasis on turn-over features and Arrhenius plot in thermal activated 
processes assisted by tunneling.


\section{A Coherent State Representation of Quantum Noise and 
Quantum Langevin Equation in c-numbers}

We consider a particle of unit mass coupled to a medium comprised 
of a set of harmonic
oscillators with frequency $\omega_i$. This is described by the following
Hamiltonian:

\begin{equation}
\label{eq1}
\hat {H} = \frac{\hat {p}^2}{2} + V(\hat{x}) + 
\sum_j \frac{\hat {p}_j^2}{2} + \frac{1}{2} \kappa_j 
({\hat{q}}_j - \hat{x})^2
\end{equation}

\noindent
Here $\hat{x}$ and $\hat{p}$ are co-ordinate and momentum operators of the
particle and the set $\{ \hat{q}_j,\hat{p}_j \}$ is the set of 
co-ordinate and momentum operators for the reservoir oscillators coupled 
linearly to the system through their coupling coefficients $\kappa_j$. 
The potential $V(\hat{x})$ is due to the external force field for the 
Brownian particle. The co-ordinate and momentum operators follow the
usual commutation relations [$\hat{x}, \hat{p}$] = $i\hbar$ and
[$\hat{q}_j, \hat{p}_j$] = $i\hbar \delta_{ij}$.
Note that in writing down the Hamiltonian no rotating 
wave approximation has been used.

Eliminating the reservoir degrees of freedom in the usual way 
\cite{whl,jkb,west} we obtain the
operator Langevin equation for the particle,
\begin{equation}
\label{eqn2}
\ddot{ \hat{x} } (t) + \int_0^t dt'  \gamma (t-t')  \dot{ \hat{x} } (t')
+ V' ( \hat{x} )
= \hat{F} (t) \; \; ,
\end{equation}

\noindent
where the noise operator $\hat{F} (t)$ and the memory kernel 
$\gamma (t)$ are given by
\begin{equation}
\label{eqn3}
\hat{F} (t) = \sum_j \left [  
\left \{ \hat{q}_j (0) - \hat{x} (0) \right \} 
\kappa_j  \cos \omega_j t +
\hat{p}_j (0)  \kappa_j^{1/2}  \sin \omega_j t  \right ]
\end{equation}

\noindent
and
\begin{equation}
\label{eqn4}
\gamma (t) = \sum_j \kappa_j  \cos \omega_j t \; \; ,
\end{equation}

\noindent
with $\kappa_j = \omega_j^2$ ( masses have been assumed to be unity ). 

The Eq.(\ref{eqn2}) is an exact quantized operator Langevin equation 
\cite{whl} for which the noise properties
of $\hat{F} (t)$ can be defined using a suitable initial canonical 
distribution of the bath co-ordinates and momenta. Our aim here is 
to replace it by an equivalent QGLE in $c$-numbers.  
To achieve this and to address the problem of quantum non-Markovian 
dynamics in terms of a 
{\it true probabilistic description} 
we proceed in two steps \cite{skb,db1,db2,db3}. 
We {\it first} carry out the 
{\it quantum mechanical average} of Eq.(\ref{eqn2})
\begin{equation}
\label{eqn5}
\langle \ddot{ \hat{x} } (t) \rangle + 
\int_0^t dt'  \gamma (t-t')  \langle \dot{ \hat{x} } (t') \rangle
+ \langle V' ( \hat{x} ) \rangle
= \langle \hat{F} (t) \rangle
\end{equation}

\noindent
where the average $\langle \ldots \rangle$
is taken over the initial product separable quantum states
of the particle and the bath oscillators at $t=0$,
$| \phi \rangle \{ | \alpha_1 \rangle | \alpha_2 \rangle \ldots
| \alpha_N \rangle \} $.
Here $| \phi \rangle$ denotes any arbitrary 
initial state of the particle and
$| \alpha_i \rangle$ corresponds to the initial
coherent state of the $i$-th bath oscillator. $|\alpha_i \rangle$
is given by 
$|\alpha_i \rangle = \exp(-|\alpha_i|^2/2) 
\sum_{n_i=0}^\infty (\alpha_i^{n_i} /\sqrt{n_i !} ) | n_i \rangle $,
$\alpha_i$ being expressed in terms of the mean values of the co-ordinate
and momentum of the $i$-th oscillator,
$\langle \hat{q}_i (0) \rangle - \langle \hat{x}(0) \rangle = 
( \sqrt{\hbar} /2\omega_i)
(\alpha_i + \alpha_i^\star )$ and
$\langle \hat{p}_i (0) \rangle = i \sqrt{\hbar\omega_i/2 }
(\alpha_i^\star - \alpha_i )$, respectively.
It is important to note that $\langle \hat{F} (t) \rangle$
of Eq.(\ref{eqn5}) is a classical-like noise term which, in general, is a
non-zero number because of the quantum mechanical averaging over the 
co-ordinate and momentum operators of the bath oscillators with respect to 
the initial coherent states and arbitrary initial state of the particle
and is given by
\begin{equation}
\label{eqn6}
\langle \hat{F} (t) \rangle = \sum_j \left [  
\left \{  \langle \hat{q}_j (0) \rangle - \langle \hat{x} (0) \rangle 
\right \}  \kappa_j  \cos \omega_j t +
\langle \hat{p}_j (0) \rangle  \kappa_j^{1/2}  \sin \omega_j t 
 \right ] \; \; .
\end{equation}

\noindent
It is convenient to rewrite the $c$-number equation (\ref{eqn5})
as follows;
\begin{equation}
\label{eqn7}
\langle \ddot{ \hat{x} } (t) \rangle + 
\int_0^t dt'  \gamma (t-t')  \langle \dot{ \hat{x} } (t') \rangle
+ \langle V' ( \hat{x} ) \rangle
= F (t)
\end{equation}

\noindent
where we let the quantum mechanical mean value 
$ \langle \hat{F} (t) \rangle \equiv F (t)$. We now turn to the {\it second}
averaging. To realize $F(t)$ as an effective 
$c$-number noise we now assume that the momenta 
$\langle \hat{p}_j (0) \rangle$ and the shifted co-ordinates
$\{ \langle \hat{q}_j (0) \rangle - \langle \hat{x} (0) \rangle \}$
of the bath oscillators are distributed according to a canonical distribution
of Gaussian forms as
\begin{equation}
\label{eqn8}
{\cal P}_j = {\cal N}
\exp \left \{ \frac{ -  [ \langle \hat{p}_j (0) \rangle^2 +
\kappa_j \left \{
\langle \hat{q}_j (0) \rangle - \langle \hat{x} (0) \rangle \right \}^2  
] }{
2 \hbar \omega_j \left ( \bar{n}_j + \frac{1}{2} \right ) }
\right \} 
\end{equation}

\noindent
so that for any quantum mechanical mean value
$O_j ( \langle\hat{p}_j (0) \rangle, 
( \langle \hat{q}_j (0) \rangle  - \langle \hat{x} (0) \rangle )$ the 
statistical average $\langle \ldots \rangle_S$ is
\begin{eqnarray}
\langle O_j \rangle_S & = & \int 
O_j ( \langle \hat{p}_j (0) \rangle, 
\{ \langle \hat{q}_j (0) \rangle  - \langle \hat{x} (0) \rangle \} )
\nonumber \\
& & \times  {\cal P}_j ( \langle \hat{p}_j (0) \rangle, 
\{ \langle \hat{q}_j (0) \rangle - \langle \hat{x} (0) \rangle \} ) 
\nonumber \\
& & \times d\langle \hat{p}_j (0) \rangle   
d \{ \langle \hat{q}_j (0) \rangle - \langle \hat{x} (0) \rangle \} \; \; .
\label{eqn9}
\end{eqnarray}

\noindent
Here $\bar{n}_j$ indicates the average thermal photon number of the $j$-th
oscillator at temperature $T$ and 
$\bar{n}_j = 1/[\exp \left ( \hbar \omega_j/k_BT \right ) - 1]$ and
${\cal N}$ is the normalization constant.

The distribution (\ref{eqn8}) and the definition of statistical average
(\ref{eqn9}) imply that $F(t)$ must satisfy

\begin{equation}
\label{eqn10}
\langle F(t) \rangle_S = 0
\end{equation}

\noindent
and

\begin{equation}
\label{eqn11}
\langle F(t)F(t^{\prime})\rangle_S = 
\frac{1}{2} \sum_j \kappa_j \hbar \omega_j
\left ( \coth \frac { \hbar \omega_j }{ 2k_BT } \right ) 
\cos \omega_j (t - t^{\prime})
\end{equation}

\noindent
That is $c$-number noise $F(t)$ is such that it is zero-centered and 
satisfies the standard fluctuation-dissipation relation (FDR) as known in 
the literature in terms of the quantum statistical average of the noise
operators. We emphasize that the method of ensemble averaging over $c$-number
noise $F(t)$ is distinctly different from traditional methods of quantum
statistical averaging and also from the method of Glauber-Sudarshan 
$P$-distribution. We discuss this in Appendix A.

We now add the force term $V^\prime(\langle\hat{x}
\rangle)$ on both sides of Eq.(\ref{eqn7}) and rearrange it to obtain

\begin{subequations}
\begin{eqnarray}
\dot X &=& P
\label{eqn12a}\\ 
\dot P &=& - \int_0^t dt^\prime \gamma (t - t^\prime) P(t^\prime) - 
V^\prime (X) + F(t) + Q(t)
\label{eqn12b}
\end{eqnarray}
\end{subequations}

\noindent
where we put $\langle\hat{x}(t)\rangle = X(t)$ for notational convenience 
and 

\begin{equation}
\label{eqn13}
Q(t) = V^\prime(\langle\hat{x}\rangle) - \langle V^\prime(\hat{x}) \rangle
\end{equation}

\noindent
represents the quantum correction due to system degrees of freedom. Since
$Q(t)$ is a fluctuation term 
Eq.(\ref{eqn12b}) offers a simple interpretation.
This implies that the QGLE is governed by a $c$-number quantum noise $F(t)$
originating from the heat bath characterized by the properties
(\ref{eqn10}) and (\ref{eqn11}) and a quantum fluctuation term characteristic
of the non-linearity of the potential. To proceed further we need a recipe 
for the calculation of $Q(t)$. This has been discussed earlier in several
contexts \cite{skb,db1,db2,db3,sm,akp}. 
For the present analysis we summarize it as follows.

Referring to the quantum nature of the system in the Heisenberg picture, one
may write.

\begin{subequations}
\begin{eqnarray}
\label{eqn14a}
\hat{x} (t) &=& \langle\hat{x}\rangle + \delta\hat{x}   \\ 
\label{eqn14b}
\hat{p} (t) &=& \langle\hat{p}\rangle + \delta\hat{p}
\end{eqnarray}
\end{subequations}

\noindent
where $\langle\hat{x}\rangle$ and $\langle\hat{p}\rangle$ are the 
quantum-mechanical averages
and $\delta\hat{x}$, $\delta\hat{p}$ are the operators. By construction 
$\langle\delta\hat{x}\rangle$ and $\langle\delta\hat{p}\rangle$ are zero
and $[\delta\hat{x},\delta\hat{p}] = i\hbar$. Using Eqs.(\ref{eqn14a}) and 
(\ref{eqn14b}) in
$\langle V^{\prime} (\hat{x}) \rangle$ and a Taylor series expansion around
$\langle\hat{x}\rangle$ it is possible to express $Q(t)$ as

\begin{equation}
\label{eqn15}
Q(t) = -\sum_{n \ge 2} \frac{1}{n!} 
V^{(n+1)} (X) \langle\delta\hat{x}^n(t)\rangle
\end{equation}

\noindent
Here $V^{n}(X)$ is the $n$-th derivative of the potential $V(X)$. To the 
second order $Q(t)$ is given by $Q(t) = -\frac{1}{2} V^{\prime\prime\prime}
(X) \langle \delta \hat{x}^2 \rangle$ where $X(t)$ and 
$\langle \delta \hat{x}^2 (t) \rangle$ can be obtained as explicit
functions of time by solving the following set of coupled equations 
(\ref{eqn16})-(\ref{eqn18}) along with Eqs.(\ref{eqn12a}) and (\ref{eqn12b});

\begin{eqnarray}
\label{eqn16}
d \langle \delta \hat{x}^2 \rangle/dt &=& 
\langle \delta \hat{x} \delta \hat{p} +
\delta \hat{p} \delta \hat{x} \rangle \\
\label{eqn17}
d \langle \delta \hat{x} \delta \hat{p} + \delta \hat{p} 
\delta \hat{x} \rangle /dt &=& 2\langle \delta \hat{p}^2 \rangle -
2V^{\prime\prime}(X) \langle \delta \hat{x}^2 \rangle  \\
\label{eqn18}
d \langle \delta \hat{p}^2 \rangle /dt  &=& -V^{\prime\prime}(X)
\langle \delta \hat{x} \delta \hat{p} + \delta \hat{p} 
\delta \hat{x} \rangle 
\end{eqnarray}

We now make some pertinent remarks on the 
above equations for quantum corrections 
and the related approximations and their successive
improvements. The set of equations 
(\ref{eqn16})-(\ref{eqn18}) for calculation of
quantum correction term to the lowest order on based on \cite{dali} the
equations $ \dot{ \delta \hat{x} } = \delta \hat{p} $ and 
$ \dot{ \delta \hat{p} } = -V^{\prime\prime}(X) \delta \hat{x} $, when the 
system-bath correlation is taken into account through $X$ in
$V^{\prime\prime}(X)$ which is obtained as a solution to Eqs.(\ref{eqn12a}) 
and (\ref{eqn12b}). The system-reservoir correlation is included at the
non-Markovian level in the dynamics involving quantum correction terms.
This implies that we have neglected the two time correlation of the system
like $ \langle \delta \hat{p} (t)  \delta \hat{p} (t^\prime) \rangle $.
The next improvement would be to include the contribution due to these
terms at least in the Markovian level to the quantum correction equations 
so that the right hand sides of Eqs.(\ref{eqn17}) and (\ref{eqn18}) may be
modified to add $ -\Gamma \langle \delta\hat{x} \delta\hat{p} +
\delta\hat{p} \delta\hat{x} \rangle $ and 
$ -2\Gamma \langle \delta\hat{p}^2 \rangle$, respectively. 
$\Gamma$ is the Markovian limit of dissipation. 
The quantum correction dynamics is thus taken into 
consideration at the Markovian level through $\Gamma$ term and at the 
non-Markovian level through $V^{\prime\prime}(X)$ term
while the mean values follow 
non-Markovian dynamics. Being exponential in nature the contribution of 
the former terms are felt only at long times when the quantum fluctuations
in the low order has less significant role to play with. Furthermore, it is
now well established \cite{bonci,els,santa,bcb} 
that curvature of the potential
$V^{\prime\prime}(X)$ controls the initial growth of the quantum 
fluctuations. Since $X(t)$ is stochastic $V^{\prime\prime}(X)$ imparts 
stochasticity in $ \ddot{ \delta \hat {x} } = 
-V^{\prime\prime}(X) \delta \hat{x} $, which makes the average dynamics of 
$\langle \delta\hat{x}^2 \rangle$ exponentially diverging 
\cite{van}. We thus emphasize that the $V^{\prime\prime}(X)$ term
dominantly controls the evolution of quantum fluctuations to a good degree
of accuracy. 
In the overdamped limit, however, it is easy to incorporate quantum 
correction to very high orders. To this end one may start from equations
$ \dot{ \delta \hat{x} } = \delta \hat{p} $ and 
$ \dot{ \delta \hat{p} } = -\Gamma \delta \hat{p} - 
V^{\prime\prime}(X) \delta \hat{x} $. Neglecting the inertial term  
$ \langle \delta \ddot{X} \rangle $ in this limit we may write 
$ -\Gamma \delta \dot {\hat{x}} = V^{\prime\prime}(X) \delta \hat{x} $ to
yield the solutions $ \langle \delta {\hat{{x}}}^n (t) \rangle $ 
for arbitrary $n$ in the expression for $Q(t)$. This has been discussed in 
greater detail in \cite{db3} in the of quantum Smoluchowskii equation.  
For a more accurate estimate of quantum corrections one may use higher order
terms in $Q(t)$ which can be calculated by solving higher order (say, twelve
equations for fourth order) coupled equations as derived in Appendix B. 
Under very special circumstance, it has been
possible to include quantum effects to all orders \cite{akp}. 

The above method of $c$-number representation of quantum Langevin
equation has been utilized recently by us to derive exact quantum Kramers'
equation \cite{db1,db2} in terms of the 
probability distribution function for calculation of 
escape rate from a metastable well and to
address several other issues. In what
follows we present a method for generation of $c$-number noise for numerical
simulation of quantum Langevin equation (\ref{eqn12a}) and (\ref{eqn12b})
in the next section.


\section{a Numerical Method For Generation of Quantum Noise}

Eq.(\ref{eqn11}), the fluctuation-dissipation relation is the key element 
for generation of quantum noise. The left hand side of this equation suggests
that $F(t)$ is a `classical' random number and $\langle F(t) F(t^\prime)
\rangle_S$ is a correlation function which is classical in form but 
quantum-mechanical in its content. That is $F(t)$ is a $c$-number Langevin
force such that Eq.(\ref{eqn12b}) (along with (\ref{eqn12a})) represents a 
stochastic differential equation. Since one of the most commonly occuring 
noise processes is exponentially correlated colour noise, we develop the 
scheme first for this process. This is based on a simple idea that $c$-number
quantum noise may be viewed as a {\it superposition} of several 
Ornstein-Ulhenbeck noise processes. 
It may be noted that Imamo\"glu \cite{ima}, 
a few years ago, has produced a scheme for 
approximating a class of reservoir spectral functions by a finite 
superposition of Lorentzian functions with positive coefficients 
corresponding to a set of fictitous harmonic oscillator modes to incorporate
non-Markovian effects into the dynamics. The aim of the work \cite{ima} is to 
develop a non-Markovian dynamics in terms of a master equation corresponding
to the system-reservoir interaction for the expanded model after the 
tracing operation. The method is an extension of the stochastic wave function 
method in quantum optics. The present superposition method 
is reminiscent of this
approach in the time domain. We however emphasize two important points. 
Because of rotating wave and Born approximations the master equation is
valid for weak system-reservoir coupling i.e., weak damping regime. The 
present method on the other hand uses a superposition of noises in the time
domain and the quantum Langevin equation 
is valid for arbitrary coupling since no weak coupling 
approximation or rotating wave approximation
has been used in the treatment. Secondly, while the 
stochastic wave function approach as well as the density operator equation
have no classical analogue, the present scheme makes use of 
$c$-numbers to implement the classical numerical method quite freely to 
solve quantum Langevin equation. 

\subsection{SUPERPOSITION METHOD}

\noindent
We proceed in several steps as follows;

\subsubsection{Step I}

\noindent
We first denote the correlation function $c(t - t^\prime)$ as

\begin{equation}
\label{eqn19}
\langle F(t)F(t^\prime) \rangle_S = c(t - t^\prime)
\end{equation}

\noindent
and in the continuum limit $c(t - t^\prime)$ is given by

\begin{equation}
\label{eqn20}
c(t - t^\prime) = \frac{1}{2} \int_0^\infty d\omega \kappa(\omega)
\rho(\omega) \hbar\omega \coth (\frac{\hbar\omega}{2k_BT}) 
\cos \omega(t - t^\prime)
\end{equation}

\noindent
where we have introduced the density function $\rho(\omega)$ for the bath 
modes. In principle an a priori knowledge of $\kappa(\omega)\rho(\omega)$
is essential and primary factor for determining
the evolution of stochastic dynamics
governed by Eqs.(\ref{eqn12a}) and (\ref{eqn12b}). We assume a Lorentzian
distribution of modes so that 

\begin{equation}
\label{eqn21}
\kappa(\omega)\rho(\omega) = \frac{2}{\pi} 
\frac{\Gamma}{1 + \omega^2\tau_c^2}
\end{equation}

\noindent
where $\Gamma$ and $\tau_c$ are the dissipation in the Markovian limit and
correlation time, respectively. These two quantities and the temperature
$T$ are the three input parameters for our problem. Eq.(\ref{eqn21}) when
used in (\ref{eqn4}) in the 
continuum limit yields an exponential memory kernel
$\gamma(t) = (\Gamma/\tau_c) \exp (-|t|/\tau_c)$. For a given set of these
parameters we first numerically evaluate the integral (\ref{eqn20}) as a
function of the time $(t - t^\prime)$. The typical profile of such a 
correlation is shown in Fig.1(a) for three different temperatures (solid
line) and for fixed $\Gamma = 1$ and $\tau_c = 1$. Similarly we present the
numerically calculated correlation functions for different values of 
$\Gamma$ for a fixed temperature 
$T = 0.0$ and correlation time $\tau_c = 1.0$;
(Fig 1(b), solid line) and for different values of 
$\tau_c$ (for fixed temperature $T = 0.0$ and
dissipation $\Gamma = 1.0$; (Fig 1(c), solid line).

\subsubsection{Step II}

In the next step we numerically {\it fit} the correlation function with a
superposition of several exponentials as

\begin{equation}
\label{eqn22}
c(t - t^\prime) = \sum_i \frac{D_i}{\tau_i} \exp 
\left( \frac{-|t - t^\prime|}{\tau_i} \right) \;\;\;\;\;\;\;\;\;\;i=1, 2, 3...
\end{equation}

\noindent
The set $D_i$ and $\tau_i$, the constant parameters are thus known. In Fig.1
(a-c) we exhibit these fitted curves (dotted lines) against the 
corresponding solid curves calculated by numerical integration of 
Eq.(\ref{eqn20}). It may be checked that at high temperatures the required
fitting curve is in general a single 
exponential, whereas at very low temperatures down
to absolute zero biexponential fitting is an excellent description. At the
intermediate temperatures one has to use even triexponential fitting, in 
general.

\subsubsection{Step-III}

\noindent
The numerical agreement between the two sets of curves based on (\ref{eqn20})
and (\ref{eqn22}) in each of the Figs.1(a-c) 
suggests that one may generate a set of 
exponentially correlated colour noise variables $\eta_i$ according to

\begin{equation}
\label{eqn23}
\dot {\eta_i} = -\frac{\eta_i}{\tau_i} + \frac{1}{\tau_i} \xi_i(t)
\end{equation}

\noindent
where

\begin{eqnarray}
\label{eqn24}
\langle \xi_i(t) \rangle &=& 0  \\
\label{eqn25}
\langle \xi_i(0) \xi_j(\tau) \rangle &=& 2D_i \delta_{ij} \delta(\tau)
\;\;\;\;\;\;\;\;\;\; (i = 1, 2, 3...)
\end{eqnarray}

\noindent
in which $\xi_i(t)$ is a Gaussian white noise obeying (\ref{eqn24}) and
(\ref{eqn25}); $\tau_i$ and $D_i$ are 
being determined in Step-II from numerical
fit. The quantum noise $\eta_i$ is thus an Ornstein-Ulhenbeck process with
properties 

\begin{eqnarray}
\label{eqn26}
\langle \eta_i(t) \rangle &=& 0  \\
\label{eqn27}
\langle \eta_i(t) \eta_j(t^\prime) \rangle &=& 
\delta_{ij} \frac{D_i}{\tau_i} 
\exp \left( \frac{-|t - t^\prime|}{\tau_i} \right)
\end{eqnarray}

\noindent
Clearly $\tau_i$ and $D_i$ are the correlation time and the strength of 
colour noise variables $\eta_i$, respectively. 
It is also important to point out that
they have nothing to do with true correlation time $\tau_c$
and strength $\Gamma$ of the noise $F(t)$ as expressed in (\ref{eqn21})
and (\ref{eqn20}).

\subsubsection{Step-IV}

\noindent
The $c$-number quantum noise $F(t)$ due to the heat bath is therefore 
given by 

\begin{equation}
\label{eqn28}
F(t) = \sum_{i=1,2..} \eta_i
\end{equation}

\noindent
Eq.(\ref{eqn28}) implies that $F(t)$ can be realized as a sum of several 
Ornstein-Ulhenbeck noises $\eta_i$ satisfying 
$\langle F(t) F(t^\prime) \rangle_S = \sum_i \langle \eta_i(t) 
\eta_i(t^\prime) \rangle$.

Having obtained the scheme for generation of $c$-number quantum noise $F(t)$
we now proceed to solve the quantum Langevin equations (\ref{eqn12a})
and (\ref{eqn12b}) along with Eqs.(\ref{eqn16})-(\ref{eqn18}). To this end
the first step is the integration \cite{fox,venk}
of the stochastic differential equation
(\ref{eqn23}). We note that to generate Gaussian white noise $\xi_i(t)$ of
(\ref{eqn23}) from two random  numbers which are uniformly distributed on
the unit interval, Box-Mueller algorithm may be used. To be specific we start
the simulation with integration of (\ref{eqn23}) to generate an initial
value of $\eta_i$ as follows;
 
\begin{eqnarray}
\label{eqn29}
m &=& {\rm random}\;\;{\rm number} \\
\label{eqn30}
n &=& {\rm random}\;\;{\rm number} \\
\label{eqn31}
\eta_i &=& \left[-\frac{2D_i}{\tau_i} \ln(m) \right]^{1/2} \cos(2\pi n)
\end{eqnarray}

\noindent
We then set $E_i = \exp(-\Delta t/\tau_i)$, where $\Delta t$ is the 
integration step length for (\ref{eqn23}). Following the integral algorithm
of Fox {\it et al} \cite{fox}, we now generate the next step 
$\eta_i|_{t + \Delta t}$
from $\eta_i$ according to the following lines;

\begin{eqnarray}
\label{eqn32}
a &=& {\rm random}\;\;{\rm number}  \\
\label{eqn33}
b &=& {\rm random}\;\;{\rm number}  \\
\label{eqn34}
h_i &=& \left[-\frac{2D_i}{\tau_i} (1 - E_i^2) 
\ln(a) \right]^{1/2} \cos(2\pi b) 
\end{eqnarray}

\noindent
and

\begin{equation}
\label{eqn35}
\eta_i|_{t + \Delta t} = \eta_i E_i + h_i
\end{equation}

\noindent
After (\ref{eqn35}) the algorithm is repeated back to (\ref{eqn32}) and
continued for a sufficient time evolution.

In the next step we couple the above-mentioned integral algorithm for
solving (\ref{eqn23}) to standard fourth order Runge-Kutta (or other)
algorithm for the solution of Eqs.(\ref{eqn12a}), (\ref{eqn12b}) and
(\ref{eqn16})-(\ref{eqn18}). For an accurate and fast algorithm 
$\eta$-integration (\ref{eqn35}) may be carried out at every $\Delta t/2$
step, while for Runge-Kutta integration of Eqs.(\ref{eqn12a}), (\ref{eqn12b}) and
(\ref{eqn16})-(\ref{eqn18}) step size $\Delta t$ is a good choice. With
$\Delta t$ in the range $\sim 0.001-0.01$ the overall method gives
excellent results for the solution of quantum Langevin equation.

\subsection{SPECTRAL METHOD}

Our method of superposition of Ornstein-Ulhenbeck noises for generation 
of quantum noise as discussed above primarily pertains to situations where
the correlation function can be described by a superposition of several
exponentials. Although in overwhelming cases of applications in condensed
matter and chemical physics at low-temperature where in general non-Ohmic 
conditions prevail we encounter such situations, there are cases where the
noise correlation exhibits power law 
behaviour or is Gaussian in nature. To extend
the present scheme we now show how the spectral method can be implemented 
for generation of quantum noise. The method had been used earlier by several
groups  \cite{stan1,stan2,rs1,rs2}
in the context of {\it classical} long and short range 
temporally correlated noises. We give here a brief outline for its extension
to quantum noise.

The starting point  of our analysis is the knowledge of quantum correlation 
function $c(t - t^\prime)$ as given in (\ref{eqn20}). Again this depends on
$\kappa(\omega) \rho(\omega)$, which (for example, for short range Gaussian
correlated noise may be taken as $2\Gamma \exp(-\omega^2\tau_c^2/4)$,
$\Gamma$ and $\tau_c$ being the strength and the 
correlation time, respectively)
must be known a priori. Having known $c(t - t^\prime)$ as a function of time
we first carry out its Fourier transform

\begin{eqnarray}
\nonumber
\langle F(t) F(t^\prime) \rangle_S &=& c(t - t^\prime) \\
\label{eqn36}
C(\omega) &=& \int\exp(-i\omega t) c(t) dt
\end{eqnarray}

\noindent
by discretizing the time in $N = 2^n$ intervals of size $\Delta t$ (such that
$\Delta t$ is shorter than any other characteristic time of the system). The
result is a string of discrete numbers $C(\omega_n)$. In discrete 
Fourier space the noise correlation is given by 

\begin{equation}
\label{eqn37}
\langle F(\omega_n) F(\omega_m^\prime) \rangle_S =
N\Delta t C(\omega_n) \delta_{m + n,0}
\end{equation}

\noindent
Therefore the noise in the Fourier domain can be constructed simply as 

\begin{eqnarray}
\label{eqn38}
F(\omega_n) = \sqrt{N\Delta t C(\omega_n)} \alpha (\omega_n) \\
\nonumber
{\rm where} \;\;\;\;\;\; i = 0, 1, 2,...,N; \;\;\;\;\; 
F(\omega_0) = F(\omega_n); \;\;\;\;\; \omega_n = \frac{2\pi n}{N\Delta t}
\end{eqnarray}

\noindent
Here $\alpha_n$ are the Gaussian complex random numbers with zero mean 
and anti-delta correlated noise, such that $\alpha_n = a_n + ib_n$; $a_n$
and $b_n$ (the real and imaginary parts of $\alpha_n$) are Gaussian random
numbers with zero mean value and variance $1/2$.

\begin{eqnarray}
\label{eqn39}  
\langle a_n \rangle &=& \langle b_n \rangle = 0; \\
\label{eqn40}
\langle a_n^2 \rangle &=& \langle b_n^2 \rangle = \frac{1}{2}; 
\;\;\;\;\; {\rm where} \;\; n \neq 0  \\
\label{eqn41}  
{\rm and}\;\;\;\; a_0^2 &=& 1
\end{eqnarray}

\noindent
Box-Mueller algorithm may again be used to generate the set of random numbers
$a_n$, $b_n$ (and consequently $\alpha_n$) according to the lines;

\begin{eqnarray}
\label{eqn42}
p_n &=& {\rm random} \;\; {\rm number}  \\
\label{eqn43}
q_n &=& {\rm random} \;\; {\rm number}   \\
\label{eqn44}
a_n &=& (-\ln{p_n})^{1/2} \cos 2\pi q_n   
\end{eqnarray}

\noindent
and similarly for $b_n$. The discrete Fourier transform of the string 
$F(\omega_n)$ is then numerically calculated by the Fast Fourier
Transform algorithm to generate the quantum random numbers $F(t_n)$. 
The recipe can be implemented for solving Eqs.(\ref{eqn12a}) and 
(\ref{eqn12b}) and (\ref{eqn16})-(\ref{eqn18}) as discussed in the earlier
subsection III-A.


\section{Application: Quantum Mean First Passage Time Calculations}

We now present the numerical results on $c$-number quantum Langevin equation
for a cubic potential. The mean first passage time whose inverse represents
the quantum rate of crossing the barrier is computed for different 
temperatures, correlation times and damping constants. 
The notion of mean in mean first passage time calculation needs some 
clarification. We are concerned here with evolution of quantum mean values
$q(=\langle \hat{q} \rangle)$ and $p(=\langle \hat{p} \rangle)$ which can be 
simultaneously determined unlike their operator counterparts 
$ \hat{q} $ and $ \hat{p} $,
respectively. The quantum fluctuation around the mean path are however
taken into account by following the equations of quantum corrections. Since
the system is coupled to heat bath it is also necessary to take statistical
`mean' over several thousands of realizations of trajectories of the quantum 
mean values.
One of the major aims of this numerical 
simulation is to show that the method described is 
subsection III-A can be implemented very easily to describe both thermally
activated processes and tunneling within an unified scheme.

To begin with we consider a potential of the form 

\begin{equation}
\label{eqn45}
V(x) = -\frac{1}{3} Ax^3 + Bx^2
\end{equation}

\noindent
where $A$ and $B$ are the constants. The other input parameters for our
calculations (Eqs.(\ref{eqn20}) and (\ref{eqn21})) are temperature $T$,
strength of noise correlation $\Gamma$ and correlation time $\tau_c$. For
the potential $V(x)$, $Q(t)$ is given by a single quantum correction term
$Q = A\langle \delta \hat{x}^2 \rangle$ according to Eq.(\ref{eqn15}).
Theoretical consideration demands that the activation barrier $\Delta E$ must
be much larger than the thermal noise, i.e, $\Delta E \gg k_BT$, where 
$\Delta E$ denotes the height of the potential barrier. In order to ensure
the stability of the algorithm we have kept $\Delta t/\tau_c \ll 1$ where
$\Delta t$ is the integration step size. 
For a given parameter set we have computed the quantum 
first passage time for the quantum Brownian particle in the cubic potential
starting from the bottom of the metastable well at $x = 0$ to reach the 
maximum at $x = 2B/A$ for a single realization of a stochastic path. 
The quantum mean first passage time (QMFPT) is 
calculated by averaging over the first passage times over $5000$ 
trajectories. We have used $\hbar = k_B = 1$ for the entire treatment.

We now present the results for the quantum mean first passage times obtained
by the simulation of the quantum damped cubic potential (\ref{eqn45}) driven
by quantum colour noise [Eqs.(\ref{eqn12a}),(\ref{eqn12b}),
(\ref{eqn16})-(\ref{eqn18}) and (\ref{eqn23})]. Our results are shown for the
rate $1/{\rm QMFPT}$ as a function of $\Gamma$ for three different 
temperatures $T = 3,1$ and $0.0$ in Fig.2(a-c). We have kept the 
parameter $\Delta E = 12$ with $A = 0.5$ and $\tau_c$ very small. 
It is observed that at $T = 3$ with increasing 
$\Gamma$, $1/{\rm QMFPT}$ undergoes a turnover from an inverse trend 
$(1/{\rm MFPT} \propto 1/\Gamma)$ to a linear behaviour
$(1/{\rm QMFPT} \propto \Gamma)$ in accordance with 
Kramers' classic result \cite{kra}
subsequently supported by many others
\cite{gro,hm,cn,tan,jkb,west}. 
The $1/{\rm QMFPT}$ has a maximum in 
the intermediate range. As the temperature is lowered to $T = 1$ and $T = 0$
the turnover persists with a little change of shape near the maxima. The 
Fig.2c represents the typical vacuum field assisted tunneling since the 
thermal activated process is almost absent at $T \sim 0$.

In Fig.3 we plot the rate $1/{\rm QMFPT}$ vs. $\tau_c$ for 
two different temperatures, with $A = 1.0$ and 
$\Delta E = 4.0$ and $\Gamma = 1.0$. It is observed
that as the temperature is lowered to $T = 0$ the decrease of $1/{\rm QMFPT}$
exhibits two distinct time scales.

In Fig.4(a-b) we exhibit Arrhenius plot, i.e, ${\rm QMFPT}^{-1}$ vs. $1/T$
for several values of $\Gamma$. We have fixed $\Delta E = 10.0$ with
$A = 1.5$ and $\tau_c = 1.0$. The physical situation corresponding to 
non-Ohmic underdamped to overdamped regions has been explored. It is 
apparent that when the temperature is high $\ln {\rm QMFPT}^{-1}$ vs. $1/T$
is typically linear corresponding to the century old Arrhenius classical 
result. However, when the temperature is lowered the quantum effects and 
$T^2$ dependence is observed. This result is in complete conformity with 
those reported earlier in connection with the theoretical as well as
experimental studies in mixed oxide systems and gas kinetics 
\cite{weiss,db1}. The variation of Arrhenius profiles with
$\Gamma$ is also indicative of the turnover phenomenon as discussed earlier.


\section{Conclusion}

In this paper we have proposed a simple numerical method for generation of
quantum noise  and for
solution of the associated generalized quantum $c$-number 
Langevin equation. The method is based primarily on two aspects. First, it
has been shown that it is possible to realize quantum noise (in a quantum
Langevin dynamics) which satisfies quantum fluctuation-dissipation relation
as $c$-numbers rather than as operators whose ordering and quantum 
statistical averaging procedure is markedly different from
those followed in the
present scheme. Second, the quantum correction terms which appear due to the
non-linearity of the potential of the system, can be generated systematically
order by order, so that the coupled set of equations for the stochastic 
dynamics of the system and the correction terms can be solved easily with
sufficient degree of accuracy. The method of simulation is equipped to deal
with arbitrary correlation time, 
temperature and damping strength. The mapping of 
quantum stochastic dynamics as a problem of classical simulation as 
presented  here allows us to implement classical methods in quantum 
simulations. The method is much simpler to handle compared to path integral
quantum Monte-Carlo as employed earlier in the problem of barrier crossing
dynamics \cite{topa}. 
We have shown that our simulation results on quantum mean first
passage time for a particle in a cubic metastable potential agree well with
the results obtained earlier. We hope that with the prescription given,
many problems where the time correlation does not follow a typical prescribed
dynamics can be simplified within the framework of this simulation.

\begin{acknowledgments}
The authors are indebted to the Council of Scientific and Industrial
Research (CSIR), Government of India, for financial support under 
Grant No. 01/(1740)/02/EMR-II.
\end{acknowledgments}


\begin{appendix}

\section{Fluctuation-Dissipation relation}

In this section we derive quantum fluctuation-dissipation relation in 
three different ways. 

\subsection*{Method-I: Quantum-Statistical Averaging}

To make the discussion self-contained we first establish that the 
quantum-statistical average value of a two-time correlation function of the 
random force operator $\hat{F} (t)$ using standard usage of 
quantum-statistical averaging $\langle...\rangle_{QS}$ can be written as

\begin{equation}
\label{eqnA1}
\frac{1}{2} \langle \hat{F}(t)\hat{F}(t^{\prime}) + 
\hat{F}(t^\prime)\hat{F}(t) \rangle_{QS} = 
\sum_j \frac{1}{2} \kappa_j \hbar \omega_j
\left ( \coth \frac { \hbar \omega_j }{ 2k_BT } \right ) 
\cos \omega_j (t - t^{\prime})
\end{equation}

\noindent
where $\hat{F}(t)$ is defined by Eq.(\ref{eqn3}).
Starting from a canonical distribution of bath oscillators at $t=0$ with 
respect to the free oscillator Hamiltonian $\hat{H}_0$, with a shift of
origin $\hat{q}_j^\prime (0) = \hat{q}_j(0) - \hat{x}(0)$, where 

\begin{equation}
\label{eqnA2}
\hat{H}_0 = \sum_j \left( \frac{\hat{p}_j^2 (0)}{2} + \frac{1}{2}
\kappa_j {\hat{q}}_j^{\prime 2} (0) \right),
\end{equation}

\noindent
we construct the expectation value of an arbitrary operator $\hat{O}$ as 

\begin{equation}
\label{eqnA3}
\langle \hat{O} \rangle_{QS} = \frac{
{\rm Tr}\; \hat{O} 
e^{-\hat{H}_0/k_BT}}{{\rm Tr}\; e^{-\hat{H}_0/k_BT}}.
\end{equation}

\noindent
As usual Trace implies both quantum mechanical averaging with respect to
number states of the oscillators 
$ \Pi_{n_1 = 0}^\infty \{ | n_1 \rangle \}
\Pi_{n_2 = 0}^\infty \{ | n_2 \rangle \}...
\Pi_{n_N = 0}^\infty \{ | n_N \rangle \} $ 
and the system state $| \phi \rangle$ 
followed by Boltzmann statistical averaging. 
Making use of the definition of the quantum statistical average
Eq.(\ref{eqnA3}) we first derive

\begin{equation}
\label{eqnA4}
\langle \hat{q}_j^\prime (0) \hat{q}_k^\prime (0) \rangle_{QS} = 
\frac{\langle \hat{p}_j(0) 
\hat{p}_k(0) \rangle_{QS}}{\omega_j^2} =
\delta_{jk} \frac{ \hbar \coth (\omega_j\hbar/2k_BT)}{2\omega_j}
\end{equation}

\noindent
and

\begin{equation}
\label{eqnA5}
\langle \hat{q}_j^\prime (0) \hat{p}_k(0) \rangle_{QS} =
-\langle \hat{p}_k(0) \hat{q}_j^\prime (0) \rangle_{QS} = \frac{1}{2}i\hbar
\delta_{jk}
\end{equation}

\noindent
Since the expectation value of an odd number of factors of 
$\hat{q}_j^\prime (0)$ and $\hat{p}_j(0)$ vanishes and the expectation value 
of an even number of factors is the sum of products of the pairwise 
expectation values with the order of the factors preserved we note that 
Gaussian property of the distribution of $\hat{q}_j^\prime (0)$ and 
$\hat{p}_j(0)$  is implied. 

The relations (\ref{eqnA4}) and (\ref{eqnA5}) directly yield (\ref{eqnA1}).
On the other hand quantum mechanical averaging with respect to the number
states results in 

\begin{equation}
\label{eqnA6}
\langle F(t) \rangle_{QS} = 0
\end{equation}

\subsection*{Method-II: Glauber-Sudarshan $P$-function}

In this method we first calculate the quantum mechanical average of
$1/2[\hat{F} (t) \hat{F} (t^\prime) + \hat{F} (t^\prime) \hat{F} (t)]$ with
respect to the coherent states of the bath oscillators
$| \alpha_1 \rangle | \alpha_2 \rangle...| \alpha_N \rangle$
and system state $| \phi \rangle$
at $t=0$ where $| \alpha_i \rangle$ has been defined in the text.
$| \alpha_i \rangle$ is the eigenstate of the annihilation operator
$\hat{a}_i$ such that $\hat{a}_i | \alpha_i \rangle =
\alpha_i | \alpha_i \rangle$ where $\hat{q}_i^\prime (0)$ and 
$\hat{p}_i (0)$ are expressed as usual as $\hat{q}_i^\prime (0) =
\sqrt{ \hbar/2\omega_i } [ \hat{a}_i^\dagger (0) + \hat{a}_i (0) ]$ and
$\hat{p}_i (0) = i\sqrt{ \hbar\omega_i/2 } [ \hat{a}_i^\dagger (0) - 
\hat{a}_i (0) ]$. We thus note that the coherent state averaging 
$\langle...\rangle$ leads to (we have discarded the cross terms like
$\alpha_j\alpha_k^*$ since they will vanish after the statistical averaging
that follows in the next step),

\begin{eqnarray}
\frac{1}{2} \langle \hat{F} (t) \hat{F} (t^\prime) + 
\hat{F} (t^\prime) \hat{F} (t) \rangle 
&=& \sum_j \left[ \kappa_j 
\frac{\hbar}{2\omega_j} \left( {\alpha_j^*}^2 + 2\alpha_j^* \alpha_j + 1 +
\alpha_j^2 \right) \cos \omega_jt \cos \omega_jt^\prime  \right. \nonumber \\
&-& \frac{\kappa_j \hbar \omega_j}{2} 
\left( {\alpha_j^*}^2 - 2\alpha_j^* \alpha_j -
1 + \alpha_j^2 \right) \sin \omega_jt \sin \omega_jt^\prime \nonumber \\
&+& \left. \frac{i\hbar}{4} \kappa_j \omega_j \left( 2{\alpha_j^*}^2 
- 2\alpha_j^2 \right) \sin \omega_j (t - t^\prime) \right].
\label{eqnA7}
\end{eqnarray}

We now introduce Glauber-Sudarshan distribution known as $P$-function for the
statistical averaging of the quantum mechanical expectation values,

\begin{eqnarray}
P_{GS} (\alpha_j(0), \alpha_j^*(0)) 
&=& N \exp \left( - \frac{ |\alpha_j |^2 }
{ \bar{n}_j } \right)    \nonumber \\
&=& N \exp \left[ - \left( \frac{ \langle \hat{p}_j (0) \rangle^2 +
\omega_j^2 \langle \hat{q}_j^\prime (0) \rangle^2 }
{ 2\hbar \omega_j \bar{n}_j } \right) \right]
\label{eqnA8}
\end{eqnarray}

\noindent
where $N$ is the normalization constant. Noting that with the distribution
(\ref{eqnA8}) the normalized statistical averages $\langle...\rangle_{GS}$ are
given by, 

\begin{equation}
\label{eqnA9}
\langle \alpha_j^2 \rangle_{GS} = \langle {\alpha_j^*}^2 \rangle_{GS} = 0
\;\;\;\;\;  {\rm and}  \;\;\;\;\; \langle \alpha_j \alpha_k^* \rangle_{GS} = 
\bar{n}_j \delta_{jk}
\end{equation}

With relations (\ref{eqnA9}), the statistical average (with respect to the
$P$-function) of the quantum mechanical mean value (\ref{eqnA7}) leads us to 

\begin{equation}
\label{eqnA10}
\frac{1}{2} \langle \langle \hat{F} (t) \hat{F} (t^\prime) + 
\hat{F} (t^\prime) \hat{F} (t) \rangle \rangle_{GS} =
\frac{1}{2}  \sum_j \kappa_j \hbar \omega_j
\left ( \coth \frac { \hbar \omega_j }{ 2k_BT } \right ) 
\cos \omega_j (t - t^{\prime})
\end{equation}

\noindent
and

\begin{equation}
\label{eqnA11}
\langle \langle \hat{F} (t) \rangle \rangle_{GS} = 0
\end{equation}

\subsection*{Method-III: Our Distribution Function}

In this method we bypass the operator ordering of $\hat{F} (t){\rm s}$.
Instead, we start from coherent state average of the noise operator
$\hat{F} (t)$, as given by (\ref{eqn6}), i.e., the $c$-number noise

\begin{equation}
\label{eqnA12}
\langle \hat{F} (t) \rangle = \sum_j \left [  
\left \{  \langle \hat{q}_j (0) \rangle - \langle \hat{x} (0) \rangle 
\right \}  \kappa_j  \cos \omega_j t +
\langle \hat{p}_j (0) \rangle  \kappa_j^{1/2}  \sin \omega_j t 
 \right ] \; \; .
\end{equation}

\noindent
and then calculate the product of two-time $c$-numbers 
$\langle \hat{F} (t) \rangle \langle \hat{F} (t^\prime) \rangle$. In our 
notation we have written

\begin{equation}
\label{eqnA13}
\langle \hat{F} (t) \rangle \langle \hat{F} (t^\prime) \rangle \equiv
F(t)F(t^\prime) 
\end{equation}

\noindent
Expressing $\langle \hat{q}_j^\prime (0)$ and $\langle \hat{p}_j (0)$
in terms of $\alpha_j$ and $\alpha_j^*$ as stated earlier, we may write 

\begin{eqnarray}
F(t)F(t^\prime) &=& \left\{ \sum_j \left( \kappa_j 
\sqrt{ \frac{\hbar}{2\omega_j} } (\alpha_j^* + \alpha_j) \cos \omega_jt +
i\sqrt{ \frac{\hbar\omega_j}{2} } \omega_j 
(\alpha_j^* - \alpha_j) \sin \omega_jt \right) \right\} \nonumber \\
&\times& \left\{ \sum_k \left( \kappa_k 
\sqrt{ \frac{\hbar}{2\omega_k} } (\alpha_k^* + \alpha_k) 
\cos \omega_kt^\prime + i\sqrt{ \frac{\hbar\omega_k}{2} } \omega_k 
(\alpha_k^* - \alpha_k) \sin \omega_kt^\prime \right) \right\}.
\label{eqnA14}
\end{eqnarray}

We now introduce our distribution function (\ref{eqn8}) for ensemble 
averaging $\langle...\rangle_S$ over the product $F(t)F(t^\prime)$ 

\begin{eqnarray}
P_j (\alpha_j(0), \alpha_j^*(0)) 
&=& N \exp \left( - \frac{ |\alpha_j |^2 }
{ \bar{n}_j + 1/2 } \right)    \nonumber \\
&=& N \exp \left[ - \left( \frac{ \langle \hat{p}_j (0) \rangle^2 +
\omega_j^2 \langle \hat{q}_j^\prime (0) \rangle^2 }
{ 2\hbar \omega_j (\bar{n}_j +1/2)} \right) \right]
\label{eqnA15}
\end{eqnarray}

\noindent
Noting that

\begin{equation}
\label{eqnA16}
\langle \alpha_j^2 \rangle_S = \langle {\alpha_j^*}^2 \rangle_S = 0
\;\;\;\;\;  {\rm and}  \;\;\;\;\; \langle \alpha_j \alpha_k^* \rangle_S = 
(\bar{n}_j + \frac{1}{2}) \delta_{jk},
\end{equation}

\noindent
we obtain on statistical averaging over $F(t)F(t^\prime)$ 

\begin{equation}
\label{eqnA17}
\langle F(t)F(t^\prime) \rangle_S =  
\frac{1}{2}  \sum_j \kappa_j \hbar \omega_j
\left ( \coth \frac { \hbar \omega_j }{ 2k_BT } \right ) 
\cos \omega_j (t - t^{\prime})
\end{equation}

\noindent
Several points are now in order. First, we note that the 
fluctuation-dissipation relation has been expressed in several different 
but equivalent forms in Eqs.(\ref{eqnA1}), (\ref{eqnA10}) and (\ref{eqnA17}) 
(or Eq.(\ref{eqn8})). Because of the prescription of operator ordering,
the left hand sides of Eqs.(\ref{eqnA1}) and (\ref{eqnA10}) do not permit
us to identify a $c$-number noise term. Clearly Eq.(\ref{eqnA17}) suits this
purpose and quantum noise as a number $F(t)$ can be identified in 
(\ref{eqnA17}) or Eq.(\ref{eqn8}). The ensemble averaging procedure of 
method-III is therefore followed in this work. Second, our distribution
(\ref{eqn8}) is distinctly different from the Glauber-Sudarshan distribution 
(\ref{eqnA8}) because of a vacuum term $1/2$ in the width of the 
distribution (\ref{eqn8}). Third, the origin of distinction of the
methods II and III lies in the fact that in the latter case we need not carry
out an operator ordering of noise as in the former case since we are 
esssentially dealing with a correlation of numbers rather than operators. 
The uncertainty principle is taken care of through the vacuum term $1/2$
in the width of the distribution (\ref{eqn8}) rather than through operator
ordering as in method-II. Fourth, the distribution (\ref{eqn8}) as well as
the Glauber-Sudarshan distribution (\ref{eqnA8}) reduces to classical 
Maxwell-Boltzmann distribution in the limit $\hbar\omega_j \ll k_BT$. Being
a quasi-classical distribution, (\ref{eqnA8}) becomes singular in the 
vacuum limit $\bar{n} \rightarrow 0$. The distribution (\ref{eqn8}) on the
other hand remains a valid non-singular distribution of quantum mean values
whose width is simply the natural width due to non-thermal 
vacuum fluctuations.

\section{Evolution Equations For Higher-Order Quantum Corrections}

\noindent
The equations for quantum corrections upto fourth order are listed below.

\noindent
Equations for the second order are:

\begin{eqnarray}
\frac{d}{dt} \langle \delta X^2 \rangle &=& \langle \delta X \delta P +
\delta P \delta X \rangle,                 \nonumber  \\
\frac{d}{dt} \langle \delta P^2 \rangle &=& 
-2\Gamma \langle \delta P^2 \rangle
-V^{\prime\prime} \langle \delta X \delta P + \delta P \delta X \rangle -      
V^{\prime\prime\prime} \langle \delta X \delta P \delta X \rangle,
\label{eqnB1}  \\
\frac{d}{dt} \langle \delta X \delta P + \delta P \delta X \rangle &=&
-\Gamma \langle \delta X \delta P + \delta P \delta X \rangle
2\langle \delta P^2 \rangle - 2V^{\prime\prime} \langle \delta X^2 \rangle -
V^{\prime\prime\prime} \langle \delta X^3 \rangle,   \nonumber
\end{eqnarray}

\noindent
Those for the third order are:

\begin{eqnarray}
\frac{d}{dt} \langle \delta X^3 \rangle &=&  
3\langle \delta X \delta P \delta X \rangle,  \nonumber \\
\frac{d}{dt} \langle \delta P^3 \rangle &=&  
-3\Gamma \langle \delta P^3 \rangle 
-3V^{\prime\prime} \langle \delta P \delta X \delta P \rangle +
V^{\prime\prime\prime} \left( \frac{3}{2} \langle \delta X^2 \rangle  
\langle \delta P^2 \rangle - \frac{3}{2} 
\langle \delta P \delta X^2 \delta P \rangle + \hbar^2 \right), \nonumber \\
\frac{d}{dt} \langle \delta X \delta P \delta X \rangle &=&
-\Gamma \langle \delta X \delta P \delta X \rangle +
2\langle \delta P \delta X \delta P \rangle - 
V^{\prime\prime} \langle \delta X^3 \rangle - 
\frac{V^{\prime\prime\prime}}{2} \left( \langle \delta X^4 \rangle - 
{\langle \delta X^2 \rangle}^2 \right),                 \label{eqnB2}  \\
\frac{d}{dt} \langle \delta P \delta X \delta P \rangle &=&
-2\Gamma \langle \delta P \delta X \delta P \rangle +
\langle \delta P^3 \rangle - 2V^{\prime\prime}
\langle \delta X \delta P \delta X \rangle  \nonumber \\
&+& \frac{V^{\prime\prime\prime}}{2}
\left( \langle \delta X^2 \rangle \langle \delta X \delta P + 
\delta P \delta X \rangle - \langle \delta X^3 \delta P + 
\delta P \delta X^3 \rangle \right),                           \nonumber
\end{eqnarray}

\noindent
And lastly, the fourth order equations are:

\begin{eqnarray}
\frac{d}{dt} \langle \delta X^4 \rangle &=& 
2\langle \delta X^3 \delta P + \delta P \delta X^3 \rangle,  \nonumber \\
\frac{d}{dt} \langle \delta P^4 \rangle &=& 
-4\Gamma \langle \delta P^4 \rangle
-2V^{\prime\prime} \langle \delta X \delta P^3 + \delta P^3 \delta X \rangle + 
2V^{\prime\prime\prime} \langle \delta X^2 \rangle \langle \delta P^3 \rangle,   
\nonumber  \\
\frac{d}{dt} \langle \delta X^3 \delta P + \delta P \delta X^3 \rangle &=&
-\Gamma \langle \delta X^3 \delta P + \delta P \delta X^3 \rangle
-2V^{\prime\prime} \langle \delta X^4 \rangle - 3\hbar^2 + 
6\langle \delta P \delta X^2 \delta P \rangle   \nonumber   \\   
&+& V^{\prime\prime\prime} 
\langle \delta X^2 \rangle \langle \delta X^3 \rangle,
\label{eqnB3}   \\
\frac{d}{dt} \langle \delta X \delta P^3 + \delta P^3 \delta X \rangle &=&
-3\Gamma \langle \delta X \delta P^3 + \delta P^3 \delta X \rangle +
2\langle \delta P^4 \rangle + 3V^{\prime\prime} (\hbar^2 - 
2\langle \delta P \delta X^2 \delta P \rangle)   \nonumber  \\
&+& 3V^{\prime\prime\prime}  
\langle \delta X^2 \rangle \langle \delta P \delta X \delta P \rangle,
\nonumber  \\
\frac{d}{dt} \langle \delta P \delta X^2 \delta P \rangle &=&
-2\Gamma \langle \delta P \delta X^2 \delta P \rangle
-V^{\prime\prime} \langle \delta X^3 \delta P + \delta P \delta X^3 \rangle +
\langle \delta P^3 \delta X + \delta X \delta P^3 \rangle    \nonumber \\
&+& V^{\prime\prime\prime} \langle \delta X^2 \rangle 
\langle \delta X \delta P \delta X \rangle. \nonumber
\end{eqnarray}
       
\noindent
The derivatives of $V(X)$, i.e., $V^{\prime\prime}$ or 
$V^{\prime\prime\prime}$ etc. in the above expressions are functions of $X$.

\end{appendix}


\newpage

\begin{figure}[h]
\vspace*{-3.0cm}
\includegraphics[width=0.7\linewidth,angle=-90]{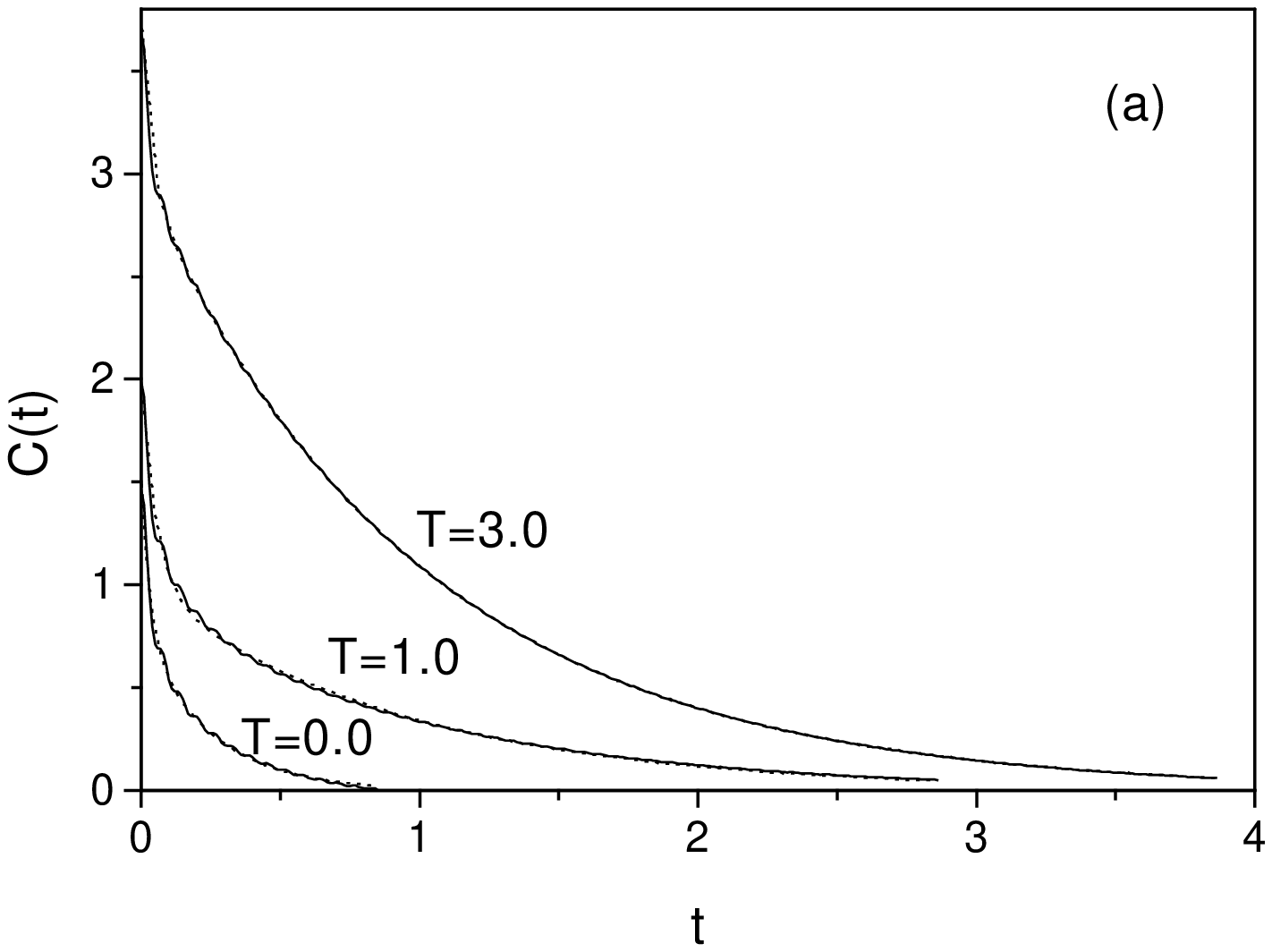}
\vspace*{-0.5cm}
\vspace*{-4.5cm}
\includegraphics[width=0.7\linewidth,angle=-90]{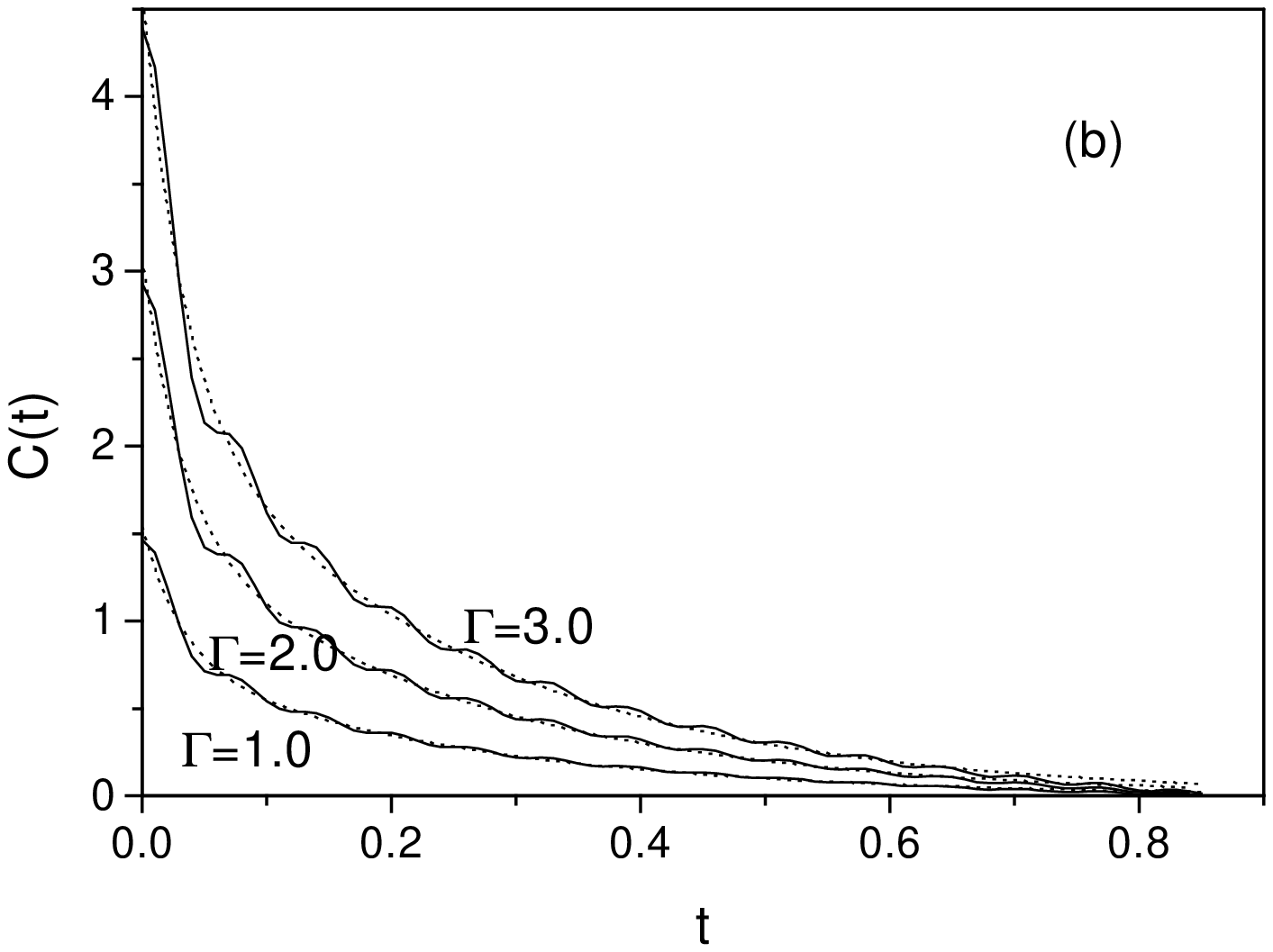}
\vspace*{-0.5cm}
\vspace*{-3.5cm}
\includegraphics[width=0.7\linewidth,angle=-90]{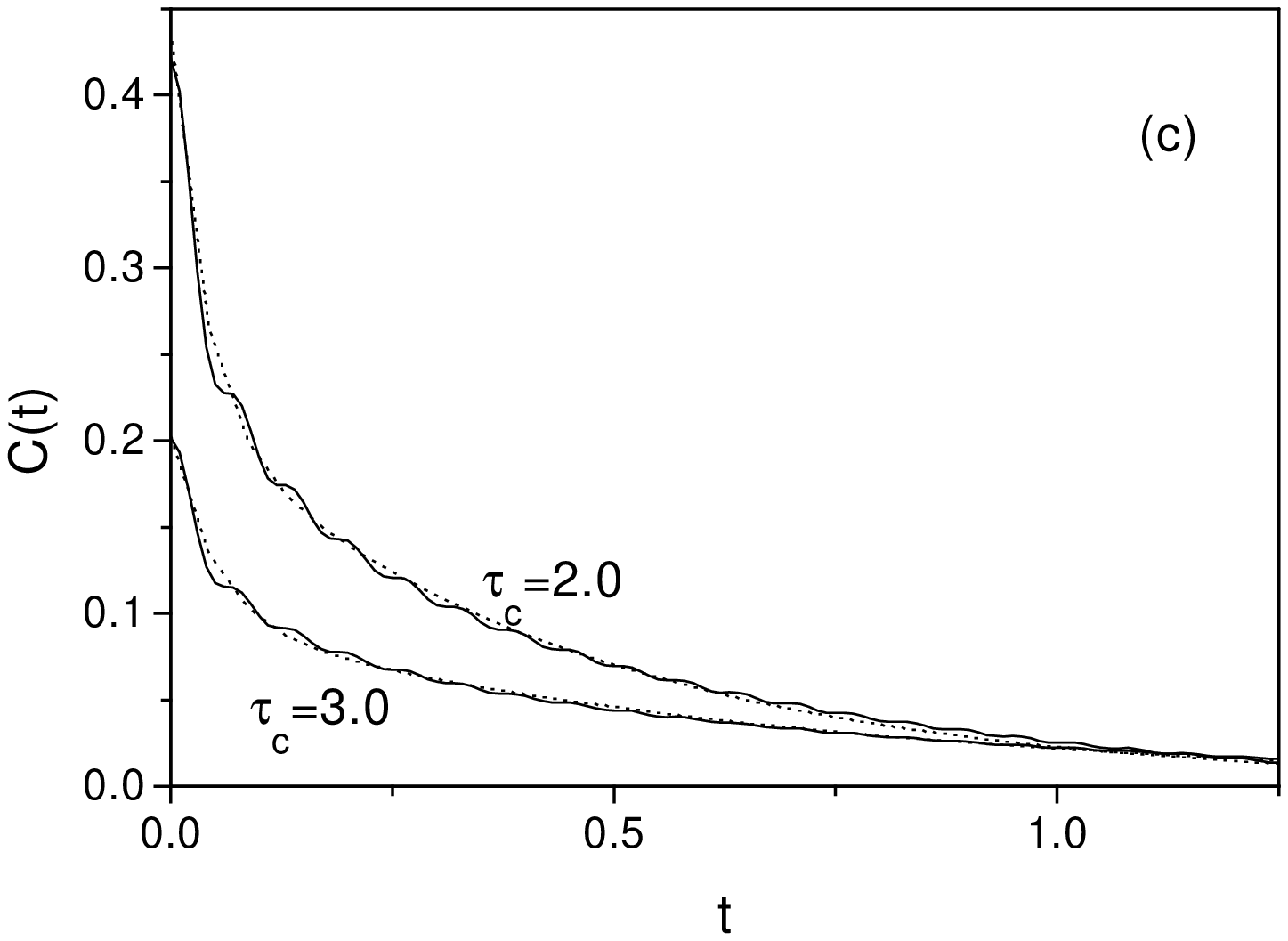}
\vspace*{-2.5cm}
\caption{
Plot of correlation function $c(t)$ vs. $t$ (units arbitrary)
as given by Eq.(20) (solid line) and Eq.(22) (dotted line) 
for the set of parameter values mentioned in the text.
}
\end{figure}

\newpage

\begin{figure}[h]
\vspace*{-3.0cm}
\includegraphics[width=0.7\linewidth,angle=-90]{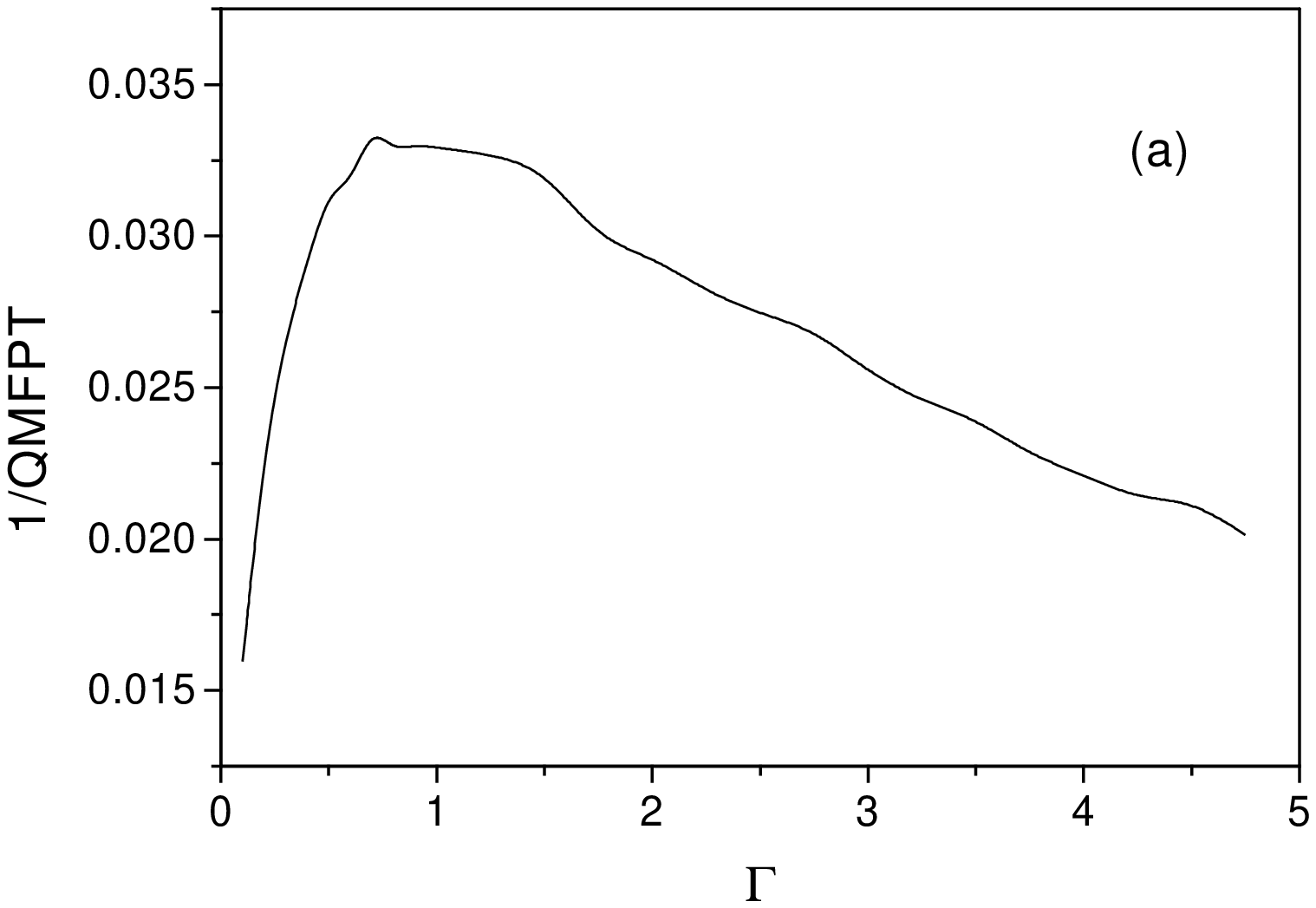}
\vspace*{-0.5cm}
\vspace*{-4.5cm}
\includegraphics[width=0.7\linewidth,angle=-90]{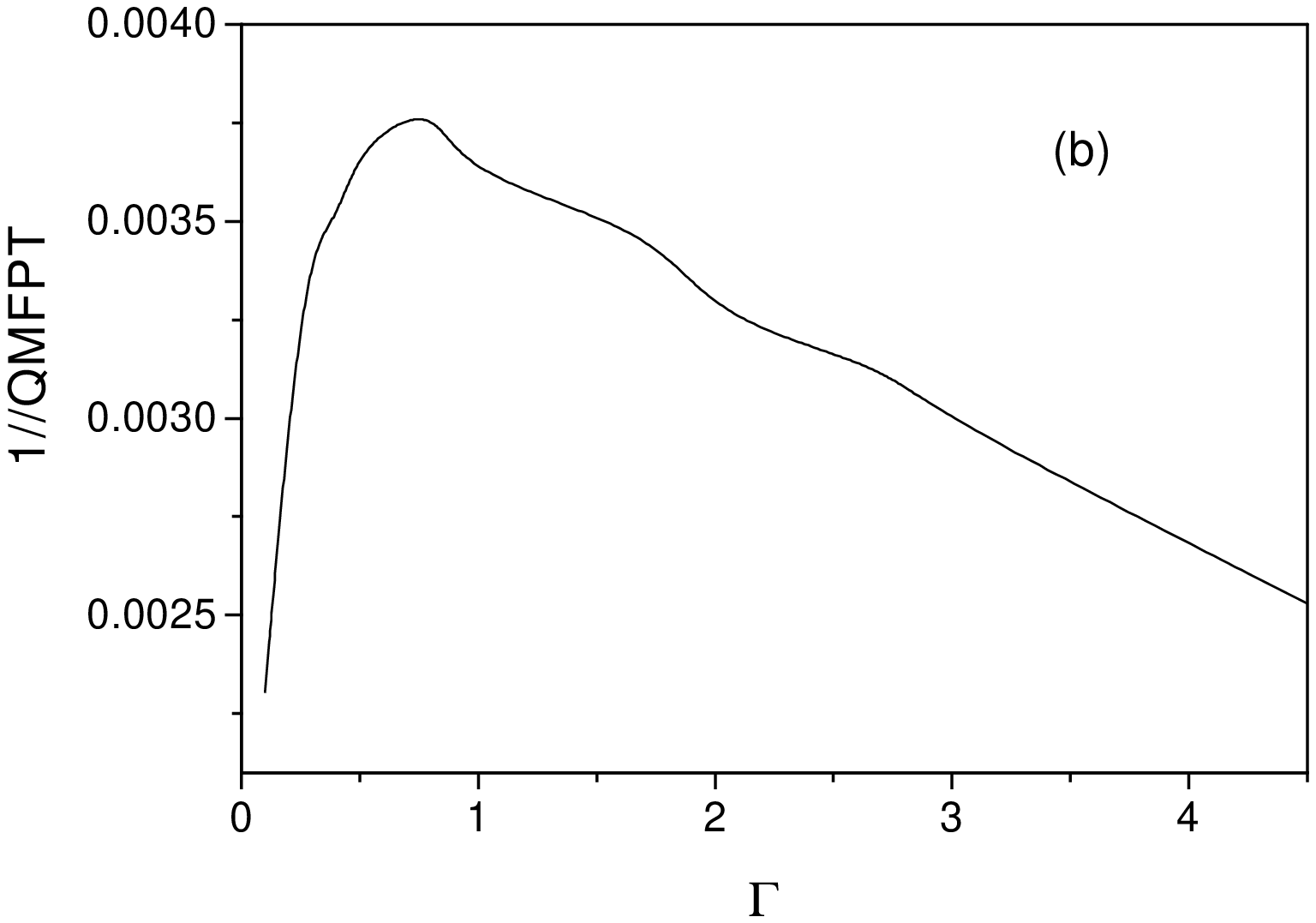}
\vspace*{-0.5cm}
\vspace*{-3.5cm}
\includegraphics[width=0.7\linewidth,angle=-90]{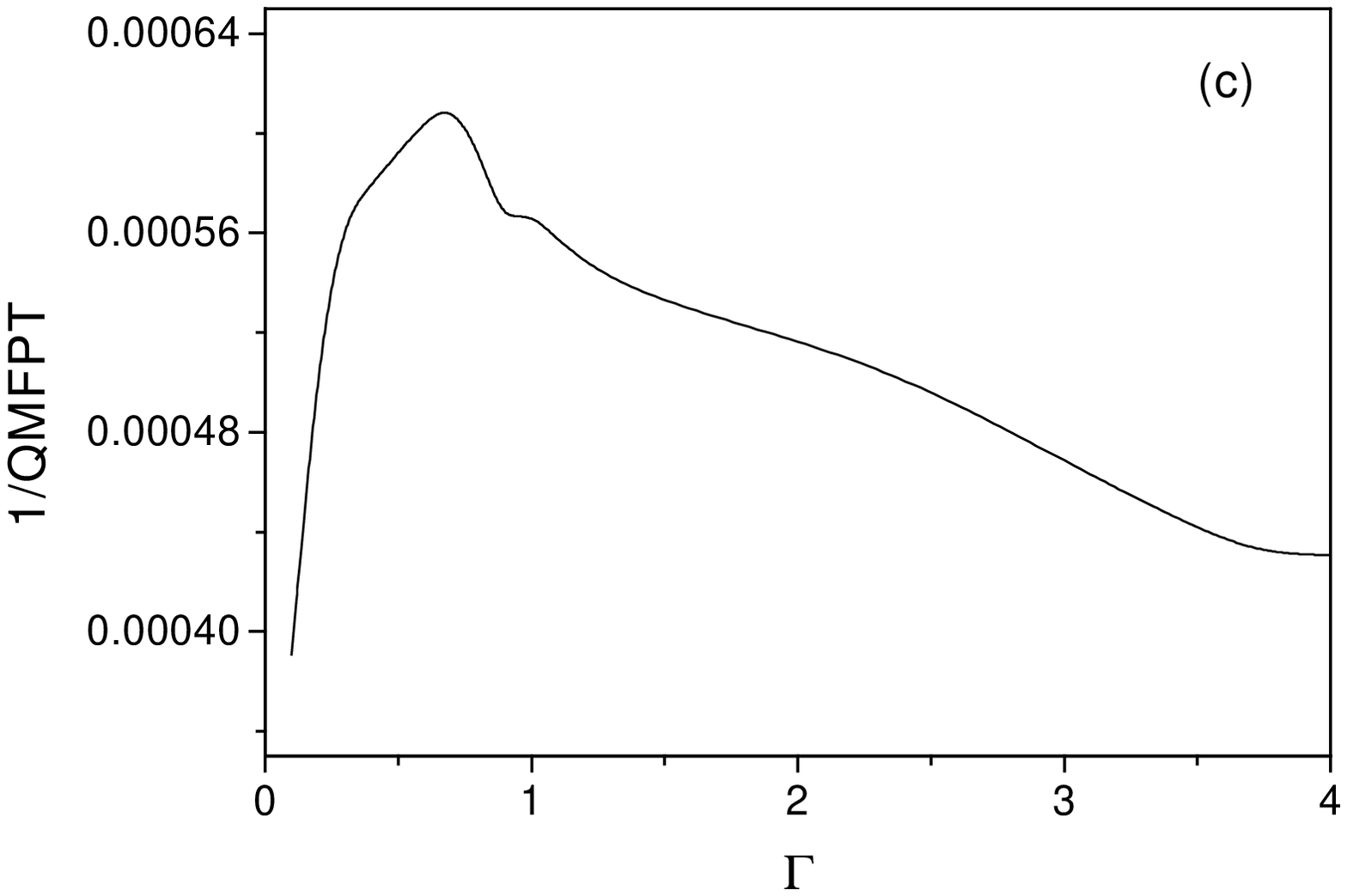}
\vspace*{-2.5cm}
\caption{
Plot of inverse of quantum mean first passage time
$\frac{1}{QMFPT}$ vs. dissipation parameter $\Gamma$ (units arbitrary)
for different temperatures: (a) $T=3.0$
(b) $T=1.0$ and (c)$T=0.0$.
}
\end{figure}

\newpage

\begin{figure}[h]
\vspace*{-3.0cm}
\includegraphics[width=0.7\linewidth,angle=-90]{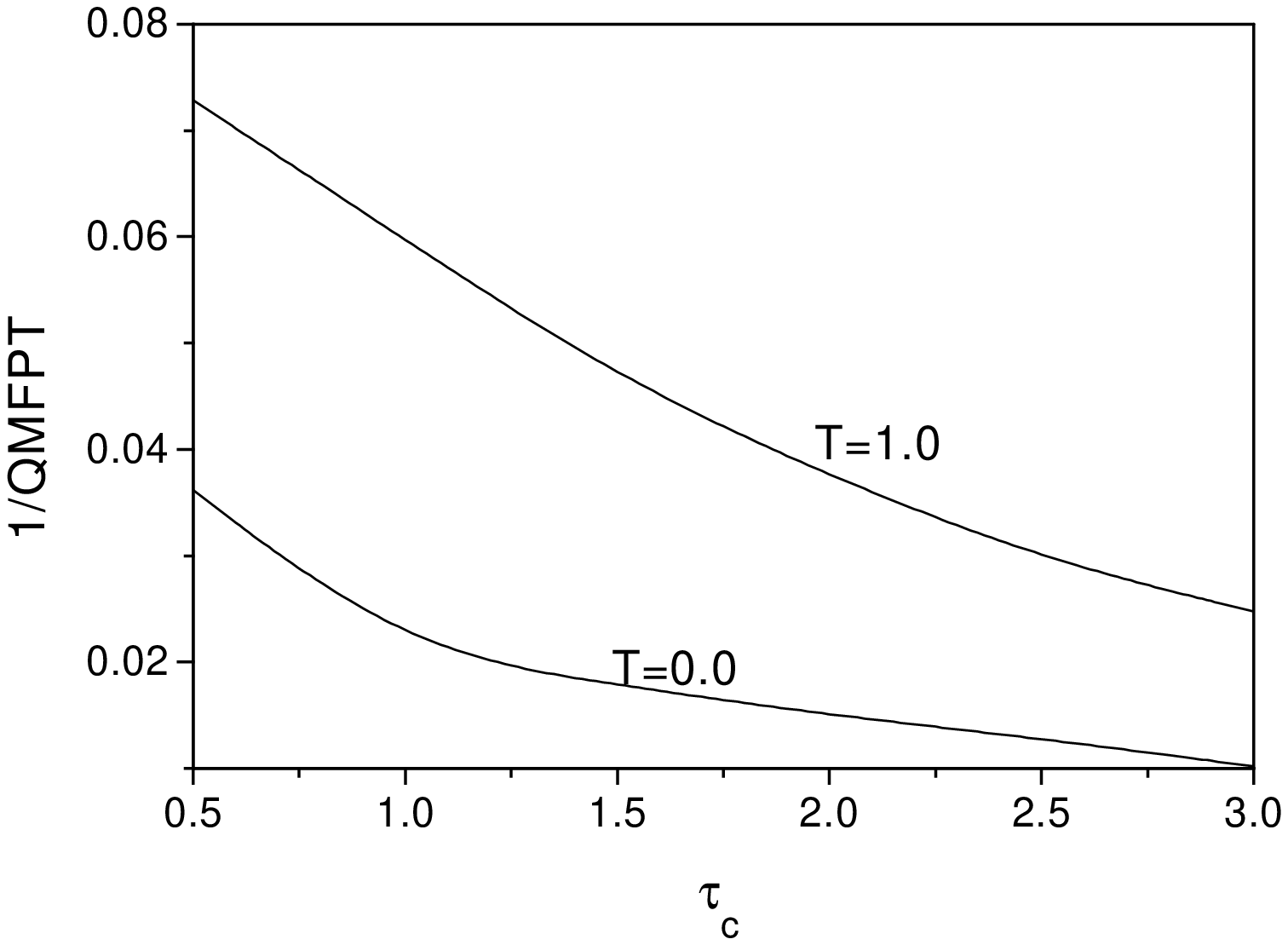}
\caption{
Plot of inverse of quantum mean first passage time
$\frac{1}{QMFPT}$ vs. correlation time  $\tau_c$ (units arbitrary) 
for two different temperatures, $T = 0.0$ and $T = 1.0$, 
and for a fixed $\Gamma = 1.0$
}
\end{figure}

\newpage

\begin{figure}[h]
\vspace*{-3.0cm}
\includegraphics[width=0.7\linewidth,angle=-90]{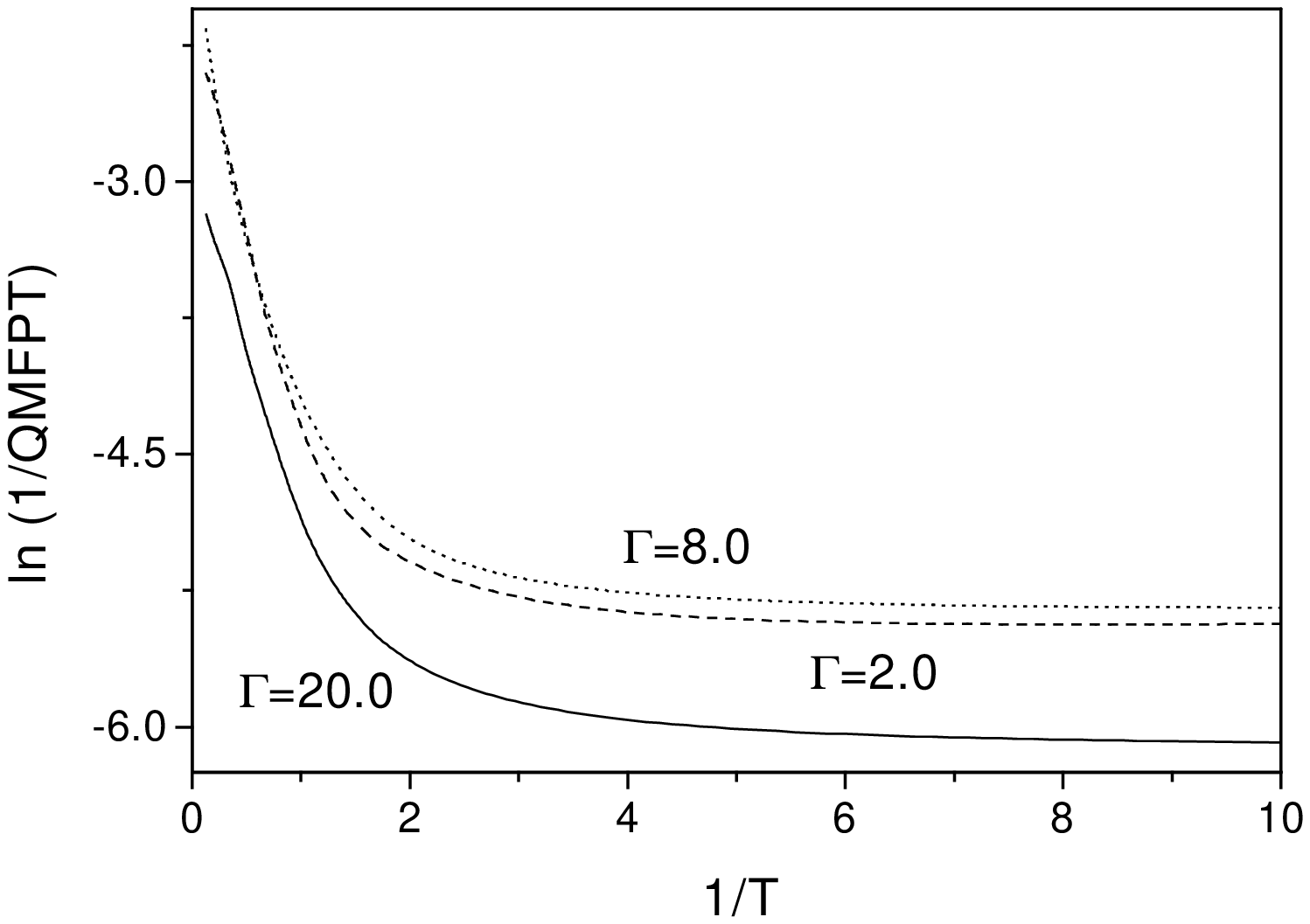}
\caption{
Plot of logarithm of inverse of quantum mean first passage time
$\ln(\frac{1}{QMFPT})$ vs. $1/T$ (units arbitrary)  
for three different $\Gamma$ values, $\Gamma = 2.0$, $\Gamma = 8.0$ and
$\Gamma = 20.0$.
}
\end{figure}

\end{document}